\documentclass[onecolumn,american]{IEEEtran}
\usepackage[T1]{fontenc}
\usepackage[utf8]{inputenc}
\usepackage{amsmath}
\usepackage{amsthm}
\usepackage{amssymb}
\usepackage{graphicx}

\makeatletter
\theoremstyle{plain}
\newtheorem{thm}{\protect\theoremname}
\theoremstyle{definition}
\newtheorem{defn}[thm]{\protect\definitionname}
\theoremstyle{remark}
\newtheorem{rem}[thm]{\protect\remarkname}
\theoremstyle{plain}
\newtheorem{cor}[thm]{\protect\corollaryname}

\makeatother

\usepackage{babel}
\providecommand{\corollaryname}{Corollary}
\providecommand{\definitionname}{Definition}
\providecommand{\remarkname}{Remark}
\providecommand{\theoremname}{Theorem}

\begin{document}
\title{One-Cold Poisson Channel: A Simple Continuous-Time Channel with Zero Dispersion}
\author{Cheuk Ting Li\\
Department of Information Engineering, The Chinese University of Hong Kong, Hong Kong, China\\
Email: ctli@ie.cuhk.edu.hk}
\maketitle
\begin{abstract}
We introduce the one-cold Poisson channel (OCPC), where the transmitter chooses one of several frequency bands to attenuate at a time. In particular, the perfect OCPC, where the number of bands is unlimited, is an extremely simple continuous-time memoryless channel. It has a capacity 1, zero channel dispersion, and an information spectrum being the degenerate distribution at 1. It is the only known nontrivial (discrete or continuous-time) memoryless channel with a closed-form formula for its optimal non-asymptotic error probability, making it the simplest channel in this sense. A potential application is optical communication with a tunable band rejection filter. Due to its simplicity, we may use it as a basic currency of information that is infinitely divisible, as an alternative to bits which are not infinitely divisible. OCPC with perfect feedback gives a generalization of prefix codes. We also study non-asymptotic coding and channel simulation results for the general OCPC.
\end{abstract}


\section{Introduction}

What is the canonical carrier of $T$ nats of information? In information theory, a binary value $X\in\{0,1\}$ is regarded as the canonical carrier of $1$ bit of information, and the \emph{universality of bits}---the fact that every information can be (approximately) converted to bits, is one of the most fundamental facts in information theory \cite{shannon1948mathematical}. Nevertheless, many questions are left unanswered. What is $0.5$ bit? What is $1$ nat $=$ $\log_{2}e$ bits? What is a natural definition of a ``canonical'' system that carries $T$ nats? There are two requirements: 1) information (like mass and length) should be infinitely divisible, so a carrier of $T_{1}+T_{2}$ nats can be divided into a carrier of $T_{1}$ nats and a carrier of $T_{2}$ nats; and 2) the carrier should consistently carry $T$ nats, not only carry $T$ nats on average. 

Discrete random variables fail these requrements, as they lack infinite divisibility \cite{li2022infinite} (we cannot divide a bit $X\sim\mathrm{Bern}(1/2)$ into two halves), and a non-uniform random variable $Y$ does not consistently carry the same amount of information (the self-information $-\log P_{Y}(Y)$ is random). For an analogy, million-dollar sports cars are not canonical carriers of monetary value, as it is impossible to have a ``half-million-dollar car'' by dividing a car into two halves. A promise to give a car if a coin flip is won is also not a ``half-million-dollar car'' since its value is random and not consistent. Hence, we focus on channels instead. Our goal is to design a channel that fulfils these requirements, and discuss its applications.

Common examples of  memoryless channels include  the binary symmetric channel (BSC), the binary erasure channel (BEC),  the additive white Gaussian noise (AWGN) channel \cite{shannon1948mathematical} and the Poisson channel \cite{kabanov1978capacity,davis1980capacity,wyner1988capacity}.  AWGN and Poisson channels are continuous-time channels that are infinitely divisible (the time duration can be arbitrarily divided), but they are noisy so the amount of information they carry is not consistent, and we have to transmit below the capacity in nonasymptotic channel coding to accommodate the variability.\footnote{One may argue that noisy channels are interesting \emph{because} of their variability, though such variability is detrimental to the use of channels as a ``basic currency'' of information. Bits can act as the universal unit of information because there is little loss in converting to/from bits, e.g., Huffman coding \cite{huffman1952method} incurs only $1$ bit of penalty.} The variability can be measured by channel dispersion \cite{hayashi2009information,polyanskiy2010channel}.  

We are particularly interested in the Poisson channel \cite{kabanov1978capacity,davis1980capacity,wyner1988capacity}, which are often used to model deep-space optical communication with photon counting receiver \cite{hemmati2006deep} and molecular communication \cite{salariseddigh2023deterministic,salariseddigh2024identification}, where the detection events form a Poisson process. One modulation scheme for the Poisson channel is pulse position modulation (PPM) \cite{hemmati2006deep,barsoum2010exit,zabini2014ppm}, where the encoder chooses one of the slots to send a pulse, so only one slot has a high level of energy. Poisson channels can be generalized to mutliple transmitters and/or multiple Poisson processes at the receiver, for example, in multiple access channels \cite{lapidoth2002poisson}, MIMO Poisson channels \cite{haas2002capacity,chakraborty2008outage}, and affine Poisson channels \cite{salariseddigh2024identification}. PPM can be combined with wavelength shift keying (WSK) which uses multiple frequencies \cite{hemmati2006deep}.

\begin{figure}
\begin{centering}
\includegraphics[scale=1.05]{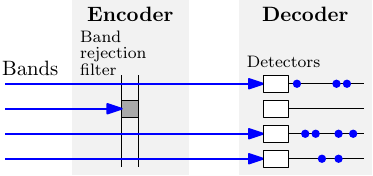}
\par\end{centering}
\caption{One-cold Poisson channel with $\mathsf{L}=4$ bands and $\alpha=0$, where the encoder blocks one band. Detection events at each band form a Poisson process with rate $1$, except the blocked band which has rate $0$.}\label{fig:ocpc}

\end{figure}

In this paper, we introduce the \emph{one-cold Poisson channel} (OCPC), where there are $\mathsf{L}$ bands (``bands'' can either be frequency bands as in WSK \cite{hemmati2006deep}, spatial diversity such as having multiple transmitters as in MIMO \cite{haas2002capacity}, polarization, types of molecules in molecular communication \cite{salariseddigh2023deterministic,salariseddigh2024identification}, or other sources of diversity), and the encoder chooses one of the bands to \emph{attenuate} by a factor $\alpha$, where the choice can change at any time. The channel input is a function $x(t)\in\{1,\ldots,\mathsf{L}\}$ which indicates the index of the band to attenuate at time $t$. The channel output consists of $\mathsf{L}$ Poisson processes $Y_{i}(t)$ for $i=1,\ldots,\mathsf{L}$, where the intensity of the process $Y_{i}(t)$ at time $t$ is $\alpha$ if $i=x(t)$, or $1$ if $i\neq x(t)$. See Figure \ref{fig:ocpc}. We will study the channel capacity, dispersion, error exponent and nonasymptotic coding results of OCPC.

If $\alpha>1$, then we amplify instead of attenuate one band, and the OCPC becomes similar to PPM and WSK. While our results also apply to the case $\alpha>1$, we are more interested in the case $0\le\alpha<1$ where we attenuate one band, similar to inverse or complementary PPM \cite{lu2012proposal,pergoloni2015merging}. Such a ``one-cold'' modulation is less common than ``one-hot'' modulation such as PPM and WSK, since having a high level of energy in all except one band is less energy-efficient for active transmitters. Nevertheless, one-cold modulation can be realistic for ``passive'' transmitters such as modulating retro-reflectors \cite{salas2012modulating}, where a satellite reflects a beam emitted from Earth unless the modulator on the satellite absorbs the beam, and the satellite communicates with Earth by choosing a small number of bands to absorb. Another application is the use of a tunable band rejection filter such as a fiber Bragg grating \cite{mohammad2004analysis} to transmit information, where a transmiter controls the strain of the grating to choose one wavelength to reject, and allow all other wavelengths to pass. One-cold modulation is desirable if only one band can be rejected (e.g., fiber Bragg grating), rejecting a band costs energy, a consistent intensity should be maintained \cite{pergoloni2015merging}, or if the beam is also used for other purposes (e.g., power transmission or satellite laser ranging) so we cannot block a significant portion of it.

An interesting special case is the \emph{perfect OCPC} where $\mathsf{L}=\infty$ and $\alpha=0$. The perfect OCPC has extremely simple behavior, considerably simpler than even the common examples of channels such as BSC, BEC and AWGN. We list some properties of the perfect OCPC:
\begin{itemize}
\item It has a capacity $1$ nat, zero channel dispersion, and an information spectrum being the degenerate distribution at $1$, meaning that it consistently carries $1$ nat per unit time.  This is the only known continuous-time memoryless channel with zero dispersion.
\item It has a simple optimal non-asymptotic channel code, with optimal error probability given exactly by
\[
1-e^{T}\mathsf{M}^{-1}\big(1-(1-e^{-T})^{\mathsf{M}}\big),
\]
where $T$ is the time duration, and $\mathsf{M}$ is the number of possible values of the message. To the best of the author's knowledge, this is the only nontrivial (discrete/continuous-time) memoryless channel with a closed-form formula for the optimal error probability.
\item The logarithm of the largest possible $\mathsf{M}$ in terms of $T$ and the error probability $0<\epsilon\le1/12$ satisfies
\begin{equation}
\ln\mathsf{M}^{*}-\max\{T-\ln(1/\epsilon),0\}\in[0,2].\label{eq:intro_M_star}
\end{equation}
Compared to most other channels (e.g., PPM Poisson channel \cite{zabini2014ppm}) where $\ln\mathsf{M}^{*}\approx CT-\sqrt{VT}Q^{-1}(\epsilon)$, $C$ is the capacity, $V$ is the channel dispersion, and $Q^{-1}$ is the inverse $Q$-function \cite{hayashi2009information,polyanskiy2010channel}, the perfect OCPC does not have the $O(\sqrt{T})$ dispersion term. The loss due to having a finite $T$ is negligible for the perfect OCPC since it has zero dispersion. 
\item The error exponent \cite{shannon1957certain} is $E(R)=1-R$ for a communication rate $0<R<1$. 
\item The identification capacity \cite{ahlswede2002identification} is infinite.
\item With perfect feedback, the optimal expected time to transmit the message $K\sim\mathrm{Unif}([\mathsf{M}])$ with zero error is the harmonic number $\sum_{m=1}^{\mathsf{M}-1}m^{-1}$. We can transmit a non-uniform source $S$ via a generalization of prefix code. 
\item Channel simulation for the perfect OCPC with duration $T$ with unlimited common randomness \cite{bennett2002entanglement,cuff2013distributed,li2024channel} can be performed using a message $K\in[\mathsf{M}]$ where $\ln\mathsf{M}\le T+\ln\ln(1/\epsilon)+\ln2$, within a total variation distance at most $\epsilon$. This shows that we can convert $T\log_{2}e+O(\ln\ln(1/\epsilon))$ bits into a perfect OCPC with duration $T$. Combined with the channel coding result (\ref{eq:intro_M_star}) which converts a perfect OCPC  into $T\log_{2}e-O(\ln(1/\epsilon))$ bits, we can convert between bits and perfect OCPC with negligible loss, showing that perfect OCPC can act as an infinitely-divisible ``basic currency'' of information.
\end{itemize}

\subsection*{Notations}

Entropy is in nats. Define $\mathbb{N}:=\{1,2,\ldots\}$, $\mathbb{Z}_{\ge0}:=\{0,1,\ldots\}$. Write $[n]:=\{1,\ldots,n\}$ and $[\infty]:=\mathbb{N}$. For an event $A$, write $\mathbf{1}\{A\}$ for the indicator of $A$ (which is $1$ if $A$ occurs, or $0$ otherwise). For a distribution $P_{X}$ and a channel $P_{Y|X}$, write $I(P_{X},P_{Y|X})$ for $I(X;Y)$ when $(X,Y)\sim P_{X}P_{Y|X}$. The information density is $\iota_{X;Y}(x;y):=\ln\frac{\mathrm{d}P_{Y|X}(\cdot|x)}{\mathrm{d}P_{Y}}(y)$. $Q^{-1}$ is the inverse $Q$-function. For two distributions $P,\tilde{P}$ over $\mathcal{X}$, the total variation distance is $d_{\mathrm{TV}}(P,\tilde{P}):=\sup_{A\subseteq\mathcal{X}\,\text{measurable}}|P(A)-\tilde{P}(A)|$.

\section{The One-Cold Poisson Channel}

We define the one-cold Poisson channel.
\begin{defn}
\label{def:ocpc}The $(\mathsf{L},\alpha)$-\emph{one-cold Poisson channel} (or $(\mathsf{L},\alpha)$-OCPC in short) with diversity parameter $\mathsf{L}\in\mathbb{N}\cup\{\infty\}$, leakage parameter $\alpha\ge0$ and duration $T>0$ has input being a measurable function $x:[0,T]\to[\mathsf{L}]$ (take $[\infty]=\mathbb{N}$), and output being a sequence of Poisson counting processes $Y_{i}(t)\in\mathbb{Z}_{\ge0}$ for $i\in[\mathsf{L}]$, $t\in[0,T]$, where $Y_{i}(t)$ has an intensity function $t\mapsto1-(1-\alpha)\mathbf{1}\{x(t)=i\}$, i.e., 
\begin{align*}
Y_{i}(t_{1})-Y_{i}(t_{0}) & \sim\mathrm{Poi}\Big(\int_{t_{0}}^{t_{1}}\big(1-(1-\alpha)\mathbf{1}\{x(\tau)=i\}\big)\mathrm{d}\tau\Big)
\end{align*}
for $0\le t_{0}<t_{1}\le T$, and is generated independently across $i$ conditional on $x$.\footnote{If we prefer the output to be a single real-valued stochastic process, we can take $Y(t)=\sum_{i=1}^{\mathsf{L}}2^{-i}Y_{i}(t)$.} In particular, when $\mathsf{L}=\infty$ and $\alpha=0$, we call it a \emph{perfect one-cold Poisson channel}.
\end{defn}
\smallskip{}

The one-cold Poisson channel is memoryless in the same sense as the Poisson channel and the continuous-time AWGN channel, that is, for any $0\le t_{0}\le t_{1}\le T$, $(Y_{i}(\tau)-Y_{i}(t_{0}))_{i\in[\mathsf{L}],\tau\in[t_{0},t_{1}]}$ depends on $x$ only through $(x(\tau))_{\tau\in[t_{0},t_{1}]}$.

An application of the OCPC is optical communication, where there are $\mathsf{L}$ frequency bands, but the transmitter is only allowed to block (if $\alpha=0$), attenuate (if $0<\alpha<1$) or amplify (if $\alpha>1$) one band at a time. Assume the transmitter chooses band $x(t)\in[\mathsf{L}]$ at time $t$. The receiver has a bandpass filter followed by a detector for each band, where the detection events for each detector follow a Poisson process. The number of detection events of band $i$ up to time $t$ is $Y_{i}(t)$. 

A special case is the $(2,0)$-OCPC, which can be regarded as a ``continuous-time BEC''. To see this, assume that the duration is divided into slots of length $\delta$, and $x(t)\in\{1,2\}$ must be kept constant within one slot. If a detection occurs in band $i$, then we know $x(t)\neq i$ in that slot, so $x(t)=3-i$. Hence, $x(t)$ cannot be recovered if no detection occurs in the slot, which happens with probability $e^{-\delta}$. Each slot is a BEC with erasure probability $e^{-\delta}$.

To study the non-asymptotic behavior of channels, we study the information spectrum \cite{han2003information} and channel dispersion \cite{hayashi2009information,polyanskiy2010channel} defined below. We allow a sequence of input distributions to accomodate the case where an optimal input distribution does not exist. Note that information spectrum and channel dispersion may not be unique \cite{polyanskiy2010channel}.

\smallskip{}

\begin{defn}
A (capacity-approaching) \emph{information spectrum} of a channel $P_{Y|X}$ with capacity $C=\sup_{P_{X}}I(P_{X},P_{Y|X})$ is a distribution $F$ over $\mathbb{R}$ such that there exists a sequence $(P_{X}^{(i)})_{i\in\mathbb{N}}$ of input distributions with $\lim_{i}I(P_{X}^{(i)},P_{Y|X})=C$ where the distribution of $\iota_{X;Y}(X;Y)$ weakly converges to $F$ when $(X,Y)\sim P_{X}^{(i)}P_{Y|X}$. A \emph{channel dispersion} of $P_{Y|X}$ is a number $V\ge0$ where there exists an information spectrum with variance $V$.
\end{defn}
\smallskip{}

When we discuss these quantities for a continuous-time memoryless channel, we fix the duration $T=1$, so the capacity and information spectrum have a unit ``nat per unit time'', and dispersion has a unit ``$\mathrm{nat}^{2}$ per unit time''. We now give the capacity, information spectrum and dispersion of OCPC. In particular, the perfect OCPC has capacity $1$, dispersion $0$, and information spectrum being the degenerate distribution at $1$, making it the simplest possible continuous-time channel in this sense.

\smallskip{}

\begin{thm}
\label{thm:capacity}The capacity of the $(\mathsf{L},\alpha)$-OCPC is
\begin{equation}
\alpha\ln\alpha-(\mathsf{L}+\alpha-1)\ln(1+(\alpha-1)/\mathsf{L}).\label{eq:capacity}
\end{equation}
An information spectrum is the distribution of
\begin{equation}
\ln\big(\alpha^{Z_{1}}(1+(\alpha-1)/\mathsf{L})^{-Z_{1}-Z_{2}}\big),\label{eq:spectrum}
\end{equation}
where $Z_{1}\sim\mathrm{Poi}(\alpha)$ is independent of $Z_{2}\sim\mathrm{Poi}(\mathsf{L}-1)$, and a channel dispersion is
\begin{equation}
\alpha\big(\ln((1+(\alpha-1)/\mathsf{L})/\alpha)\big)^{2}+(\mathsf{L}-1)\big(\ln(1+(\alpha-1)/\mathsf{L})\big)^{2}.\label{eq:dispersion}
\end{equation}
In particular, for $\mathsf{L}=\infty$, the capacity is $\alpha\ln(\alpha/e)+1$, an information spectrum is the distribution of $Z_{1}\ln\alpha-\alpha+1$ (degenerates to $1$ if $\alpha=0$), and a dispersion is $\alpha(\ln\alpha)^{2}$. 
\end{thm}
\smallskip{}

\begin{IEEEproof}
Fix any $P_{X}$. Let $\bar{Y}=(\bar{Y}_{i}(t))_{i\in[\mathsf{L}],t\in[0,T]}$ be i.i.d. Poisson processes with rate $\gamma=1+(\alpha-1)/\mathsf{L}$ for $i\in[\mathsf{L}]$. 
\begin{align*}
I(X;Y) & \le\mathbb{E}[D(P_{Y|X}(\cdot|X)\Vert P_{\bar{Y}})]\\
 & =\mathbb{E}\Big[\sum_{i=1}^{\mathsf{L}}\int_{0}^{T}\Big((1-(1-\alpha)\mathbf{1}\{X(t)=i\})\\
 & \quad\quad\cdot\ln\frac{1-(1-\alpha)\mathbf{1}\{X(t)=i\}}{e\gamma}+\gamma\Big)\mathrm{d}t\Big]\\
 & =\mathbb{E}\Big[\sum_{i=1}^{\mathsf{L}}\int_{0}^{T}\Big(\Big(\alpha\ln\frac{\alpha}{e\gamma}+\gamma\Big)\mathbf{1}\{X(t)=i\}\\
 & \quad\quad+\Big(\ln\frac{1}{e\gamma}+\gamma\Big)\mathbf{1}\{X(t)\neq i\}\Big)\mathrm{d}t\Big]\\
 & =T(\alpha\ln(\alpha e^{-1}\gamma^{-1})+\gamma+(\mathsf{L}-1)(-\ln(e\gamma)+\gamma))\\
 & =T(\alpha\ln\alpha-(\mathsf{L}+\alpha-1)\ln(1+(\alpha-1)/\mathsf{L})).
\end{align*}
The above bound is $\alpha\ln(\alpha/e)+1$ when $\mathsf{L}=\infty$. We now show how this bound can be approached for the case $\mathsf{L}<\infty$. Consider a discretized version of the OCPC with input $\tilde{\mathbf{X}}:=(\tilde{X}_{1},\ldots,\tilde{X}_{n})\in[\mathsf{L}]^{n}$, where $\tilde{\mathbf{x}}$ is mapped to the input $t\mapsto\tilde{x}_{\lceil nt/T\rceil}$ of the OCPC. Let $\tilde{Y}_{j,i}:=Y_{i}(Tj/n)-Y_{i}(T(j-1)/n)$, $\tilde{Y}_{j}:=(\tilde{Y}_{j,i})_{i\in[\mathsf{L}]}$, $\tilde{\mathbf{Y}}:=(\tilde{Y}_{j})_{j\in[n]}$. Then $\tilde{X}_{j}\to\tilde{Y}_{j}$ forms a memoryless channel, with information density (assuming $\tilde{X}_{j}\sim\mathrm{Unif}([\mathsf{L}])$ i.i.d.) 
\[
\iota(\tilde{X}_{j};\tilde{Y}_{j})=\ln\frac{P(\tilde{Y}_{j}|\tilde{X}_{j})}{P(\tilde{Y}_{j})}=\ln\frac{\alpha^{\tilde{Y}_{j,\tilde{X}_{j}}}}{\mathsf{L}^{-1}\sum_{x=1}^{\mathsf{L}}\alpha^{\tilde{Y}_{j,x}}}.
\]
If $\sum_{x}\tilde{Y}_{j,x}=0$, then $\iota(\tilde{X}_{j};\tilde{Y}_{j})=0$. If $\sum_{x}\tilde{Y}_{j,x}=1$, then $\iota(\tilde{X}_{j};\tilde{Y}_{j})=\ln(\alpha^{\tilde{Y}_{j,\tilde{X}_{j}}}/(1+(\alpha-1)/\mathsf{L}))$. Since $\mathbb{P}(\max_{j}\sum_{x}\tilde{Y}_{j,x}\le1)\to1$ as $n\to\infty$, the distribution of $\iota(\tilde{\mathbf{X}};\tilde{\mathbf{Y}})=\sum_{j}\iota(\tilde{X}_{j};\tilde{Y}_{j})$ approaches the distribution of $\ln(\alpha^{Z_{1}}(1+(\alpha-1)/\mathsf{L})^{-Z_{1}-Z_{2}})$ where $Z_{1}\sim\mathrm{Poi}(\alpha T)$ is independent of $Z_{2}\sim\mathrm{Poi}((\mathsf{L}-1)T)$, which has a mean that equals the capacity and a variance given by (\ref{eq:dispersion}) when $T=1$. The case $\mathsf{L}=\infty$ is proved by taking $\mathsf{L}\to\infty$.
\end{IEEEproof}
\smallskip{}

We now study non-asymptotic channel coding for OCPC. For the $(\mathsf{L},\alpha)$-OCPC with duration $T$, let $\mathsf{M}^{*}(\mathsf{L},\alpha,T,\epsilon)$ be the largest possible $\mathsf{M}\in\mathbb{N}$ such that there exists a coding scheme for sending a message $K\sim\mathrm{Unif}([\mathsf{M}])$ with average error probability $\le\epsilon$, and let $\epsilon^{*}(\mathsf{L},\alpha,T,\mathsf{M}):=\inf\{\epsilon\ge0:\mathsf{M}^{*}(\mathsf{L},\alpha,T,\epsilon)\ge\mathsf{M}\}$ be the optimal error probability for a given $\mathsf{M}$. The following bound for finite $\mathsf{L}$ follows  from Theorem \ref{thm:capacity} and \cite[Theorem 1]{li2021unified}. The case $\mathsf{L}=\infty$ will be discussed in Section \ref{sec:sufficient_L}.

\smallskip{}

\begin{thm}
\label{thm:finite_L}For the $(\mathsf{L},\alpha)$-OCPC ($\mathsf{L}<\infty$), the optimal error probability is bounded by
\begin{align*}
 & \epsilon^{*}(\mathsf{L},\alpha,T,\mathsf{M})\;\le\\
 & \,\mathbb{E}\big[1-\big(1-\min\big\{\alpha^{-Z_{1}}(1+(\alpha-1)/\mathsf{L})^{Z_{1}+Z_{2}},1\big\}\big)^{(\mathsf{M}+1)/2}\big],
\end{align*}
where $Z_{1}\sim\mathrm{Poi}(\alpha T)$ is independent of $Z_{2}\sim\mathrm{Poi}((\mathsf{L}-1)T)$. As a result, for any fixed $0<\epsilon<1$, for $T\to\infty$,
\begin{equation}
\ln\mathsf{M}^{*}(\mathsf{L},\alpha,T,\epsilon)\ge CT-\sqrt{VT}Q^{-1}(\epsilon)-O(\log T),\label{eq:limited_L_M_star_bound}
\end{equation}
where $C$, $V$ are the capacity and dispersion in Theorem \ref{thm:capacity}.
\end{thm}
\smallskip{}

\begin{IEEEproof}
The result follows from \cite[Theorem 1]{li2021unified} which upper-bounds the error probability for a channel $P_{Y|X}$ with input distribution $P_{X}$ by $\mathbb{E}[1-(1-\min\{e^{-\iota(X;Y)},1\})^{(\mathsf{M}+1)/2}]$.  (\ref{eq:limited_L_M_star_bound}) follows from the Gaussian approximation of (\ref{eq:spectrum}) via the Berry-Esseen theorem \cite{berry1941accuracy,esseen1942liapunov}.
\end{IEEEproof}
\smallskip{}

\begin{rem}
Another way to understand the perfect OCPC is to treat it as the limit of the discrete memoryless channel $P_{Y|X}$ where $X\in[b]$ and $Y\subseteq[b]$ is a uniformly randomly chosen subset of size $a$ that contains $X$, and $1\le a\le b$.\footnote{This is the opposite of the channel in \cite[Section 3.7]{li2024channel}, and is related to the source coding setting in \cite{korner1971coding,li2025coding} where the reconstruction is a subset that contains the source.} The perfect OCPC with duration $T$ can be regarded as the limit of this channel with $a=\lceil e^{-T}b\rceil$, $b\to\infty$.
\end{rem}
\smallskip{}

\begin{rem}
We can define the OCPC with non-integer diversity $\mathsf{L}\ge1$. The \emph{fractional OCPC} has input being a measurable set $x\subseteq[0,\mathsf{L})\times[0,T]$ where the Lebesgue measure $\lambda(\{y:(y,t)\in x\})=1$ for every $t\in[0,T]$, and output being a Poisson point process $Y$ over $[0,\mathsf{L})\times[0,T]$ with intensity measure $\nu(A):=\lambda(A)-(1-\alpha)\lambda(A\cap x)$ for $A\subseteq[0,\mathsf{L})\times[0,T]$. While the fractional OCPC is more general, we focus on the OCPC in Definition \ref{def:ocpc} for simplicity. Most results in this paper (e.g., Theorem \ref{thm:sufficient_L}) also apply to fractional OCPC.
\end{rem}

\section{Optimal Non-Asymptotic Coding with Sufficient Diversity}\label{sec:sufficient_L}

We have studied non-asymptotic channel coding for OCPC with finite $\mathsf{L}$ in Theorem \ref{thm:finite_L}. Interestingly, for OCPC with sufficient diversity (i.e., $\mathsf{L}$ is infinite or no smaller than the number of possible values $\mathsf{M}$ of the message), the optimal coding scheme and error probability can be given exactly. 

\smallskip{}

\begin{thm}
\label{thm:sufficient_L}For the $(\mathsf{L},\alpha)$-OCPC with $\mathsf{L}\ge\mathsf{M}$ and $\alpha\in[0,1]$, we have $\epsilon^{*}(\mathsf{L},\alpha,T,\mathsf{M})=\mathbb{P}(\hat{K}\neq1)$, where $Z_{1}\sim\mathrm{Poi}(\alpha T)$ is independent of $Z_{i}\sim\mathrm{Poi}(T)$ i.i.d. for $i=2,\ldots,\mathsf{M}$, and $\hat{K}|(Z_{i})_{i\in[\mathsf{M}]}\sim\mathrm{Unif}(\mathrm{argmin}_{i}Z_{i})$, i.e., $\hat{K}\in[\mathsf{M}]$ is the index of the smallest $Z_{i}$, chosen randomly in case of ties. For $\alpha>1$, replace $\mathrm{argmin}$ by $\mathrm{argmax}$. In particular, for the perfect OCPC, the optimal error probability is
\[
\epsilon^{*}(\infty,0,T,\mathsf{M})=1-e^{T}\mathsf{M}^{-1}\big(1-(1-e^{-T})^{\mathsf{M}}\big).
\]
\end{thm}
\smallskip{}

 We characterize $\mathsf{M}^{*}(\infty,0,T,\epsilon)$ in Corollary \ref{cor:perfect}. Compared to most channels where $\ln\mathsf{M}^{*}\approx CT-\sqrt{VT}Q^{-1}(\epsilon)$, where $C$ is the capacity and $V$ is the channel dispersion \cite{polyanskiy2010channel,hayashi2009information}, the perfect one-cold Poisson channel has no channel dispersion. 

\smallskip{}

\begin{cor}
\label{cor:perfect}For the perfect OCPC, for $0<\epsilon\le1/12$, 
\[
\ln\mathsf{M}^{*}(\infty,0,T,\epsilon)-\max\{T-\ln(1/\epsilon),0\}\in[0,2].
\]
 For a rate $0<R<1$, the error exponent is
\[
\mathrm{liminf}_{T\to\infty}T^{-1}\ln\big(1/\epsilon^{*}(\infty,0,T,\lceil e^{TR}\rceil)\big)=1-R.
\]
\end{cor}
\smallskip{}

We now prove Theorem \ref{thm:sufficient_L} and Corollary \ref{cor:perfect}.
\begin{IEEEproof}
Consider an encoding function $x:[\mathsf{M}]\times[0,T]\to[\mathsf{L}]$, i.e., the channel input is $t\mapsto x(k,t)$ if the message is $k$. Let $U_{k}:=\{(i,t):\,x(k,t)=i\}\subseteq[\mathsf{L}]\times[0,T]$. For nonempty $S\subseteq[\mathsf{M}]$, let $W_{S}:=(\bigcap_{k\in S}U_{k})\backslash(\bigcup_{k\in[\mathsf{M}]\backslash S}U_{k})$, which form a partition of $\bigcup_{k=1}^{\mathsf{M}}U_{k}$. Intuitively, $W_{S}$ are the cells of the Venn diagram formed by the sets $U_{k}$. For measurable $A\subseteq[\mathsf{L}]\times[0,T]$, let $N(A)$ be the number of points in among $(Y_{i}(t))_{i,t}$ that lies in $A$, i.e., $N(A):=\sum_{i=1}^{\mathsf{L}}\int_{0}^{T}\mathbf{1}\{(i,t)\in A\}\mathrm{d}Y_{i}(t)$. Conditional on $K=k$, we know that 
\[
N(W_{S})\sim\mathrm{Poi}\big((1-(1-\alpha)\mathbf{1}\{k\in S\})\lambda(W_{S})\big),
\]
independent across nonempty $S\subseteq[\mathsf{M}]$, where $\lambda$ denotes the measure over $[\mathsf{L}]\times[0,T]$ that is the product of the counting measure and the Lebesgue measure. Note that $(N(W_{S}))_{S\in2^{[\mathsf{M}]}\backslash\{\emptyset\}}$ is a sufficient statistic for $K$. Given the observations $N(W_{S})=n_{S}$ for nonempty $S\subseteq[\mathsf{M}]$, the likelihood of $k$ is
\begin{align*}
 & \prod_{S\in2^{[\mathsf{M}]}\backslash\{\emptyset\}}\big((1-(1-\alpha)\mathbf{1}\{k\in S\})\lambda(W_{S})\big)^{n_{S}}\\
 & \propto\prod_{S\in2^{[\mathsf{M}]}\backslash\{\emptyset\}}(1-(1-\alpha)\mathbf{1}\{k\in S\})^{n_{S}}\,=\,\alpha^{\sum_{S\ni k}n_{S}}.
\end{align*}
Therefore, given the encoding function $x$, the optimal decoding function is $\hat{K}:=\mathrm{argmin}_{k\in[\mathsf{M}]}\sum_{S\ni k}N(W_{S})=\mathrm{argmin}_{k\in[\mathsf{M}]}N(U_{k})$ for $\alpha\le1$, or $\hat{K}:=\mathrm{argmax}_{k\in[\mathsf{M}]}N(U_{k})$ for $\alpha>1$. Ties can be broken arbitrarily, so we assume that a smaller $k$ is preferred in case of ties hereafter.

We now argue that the regions $W_{S}$ where $|S|\ge2$ (the regions that are used by more than one $k$'s) can be removed without increasing the error probability, and hence an optimal code has disjoint $U_{k}$'s. Consider any $S'$ with $|S'|\ge2$ where $\lambda(W_{S'})>0$. Construct a new code $\tilde{x}:[\mathsf{M}]\times[0,T]\to[\mathsf{L}]$ where $\tilde{x}(k,t)=x(k,t)$ if $(x(k,t),t)\notin W_{S'}$, or otherwise the $\tilde{x}(k,t)$'s are chosen such that $\tilde{x}(k,t)\in[\mathsf{L}]\backslash\{x(k,t):k\notin S'\}$ and $(\tilde{x}(k,t))_{k\in S'}$ are distinct. We can see that the new code corresponds to the sets $\tilde{W}_{S}$ where $\tilde{W}_{S'}=\emptyset$, $\tilde{W}_{\{k\}}\supseteq W_{\{k\}}$ with $\lambda(\tilde{W}_{\{k\}}\backslash W_{\{k\}})=\lambda(W_{S'})$ for $k\in S'$, and $\tilde{W}_{S}=W_{S}$ for $S\notin\{S'\}\cup\bigcup_{k\in S'}\{\{k\}\}$. Intuitively, the new code is constructed by eliminating $\tilde{W}_{S'}$, and distributing its mass to $\tilde{W}_{\{k\}}$ for $k\in S'$. We will show that the new code with a suboptimal decoder is as good as the old code with the optimal decoder. Assume $\alpha\le1$. Consider a suboptimal decoder $\tilde{K}:=\mathrm{argmin}_{k\in[\mathsf{M}]}\sum_{S\ni k}\tilde{N}_{S}$ for the new code, where $\tilde{N}_{S}=N(W_{S})$ for $S\neq S'$, and $\tilde{N}_{S'}=N(\tilde{W}_{\{K'\}}\backslash W_{\{K'\}})$ where $K':=\mathrm{argmin}_{k\in S'}\sum_{S\ni k,\,S\neq S'}N(W_{S})$. For $\alpha>1$, replace the argmins by argmaxes. Intuitively, the suboptimal decoder is the same as the optimal decoder for the old code, except that $N(W_{S'})$ is replaced by $N(\tilde{W}_{\{K'\}}\backslash W_{\{K'\}})$ where $K'\in S'$ is the ``most promising'' value among $S'$. 

We now show that the suboptimal decoder is as good as the the old code with the optimal decoder. For the case where the transmitted message is $k_{0}\notin S'$, i.e., the rate of points in $U_{k_{0}}$ is changed from $1$ to $\alpha$, $\tilde{N}_{S'}\sim\mathrm{Poi}(\lambda(W_{S'}))$ is independent of $(\tilde{N}_{S})_{S\neq S'}$ since $\tilde{W}_{\{k\}}\backslash W_{\{k\}}$ for $k\in S'$ has no overlapping with $U_{k_{0}}$. Hence, the joint distribution of $(\tilde{N}_{S})_{S}$ is the same as that of $(N(W_{S}))_{S}$. Consider the case where the transmitted message is $k_{0}\in S'$, and let $\bar{K}:=\mathrm{argmin}_{k\in[\mathsf{M}]}\sum_{S\ni k}\bar{N}_{S}$ where $\bar{N}_{S}=N(W_{S})$ for $S\neq S'$, and $\bar{N}_{S'}=N(\tilde{W}_{\{k_{0}\}}\backslash W_{\{k_{0}\}})$. Note that $\bar{K}$ is not a proper decoder since it depends on $k_{0}$, though the distribution of $\bar{K}$ is the same as that of $\hat{K}$ of the old code since $(\bar{N}_{S})_{S}$ has the same distribution as $(N(W_{S}))_{S}$. If $k_{0}=K'$, then $\tilde{K}=\bar{K}$ by definition. If $k_{0}\neq K'$, then $k_{0}\neq\mathrm{argmin}_{k\in S'}\sum_{S\ni k}\bar{N}_{S}$, and $\bar{K}\neq k_{0}$. Hence, $\bar{K}=k_{0}$ implies $\bar{K}=K'=k_{0}$, and 
\begin{align*}
\mathbb{P}(\tilde{K}=k_{0}) & \ge\mathbb{P}(\tilde{K}=K'=k_{0})\,=\,\mathbb{P}(\bar{K}=K'=k_{0})\\
 & \;=\mathbb{P}(\bar{K}=k_{0})\,=\,\mathbb{P}(\hat{K}=k_{0}).
\end{align*}
We can see that in both cases, the suboptimal decoder for the new code has an error probability not larger than the old code with the optimal decoder. By repeating this operation, we can eliminate all regions $W_{S}$ where $|S|\ge2$, so an optimal code is simply $x(k,t)=k$, where $N(W_{\{k\}})\sim\mathrm{Poi}(\alpha T)$ if $k=k_{0}$ is the transmitted message, or $N(W_{\{k\}})\sim\mathrm{Poi}(T)$ if $k\neq k_{0}$. In case if $\alpha=0$, the correct probability is
\begin{align*}
 & \mathbb{E}\left[1/(|\{i\in\{2,\ldots,\mathsf{M}\}:\,Z_{i}=0\}|+1)\right]\\
 & =\sum_{m=0}^{\mathsf{M}-1}\binom{\mathsf{M}-1}{m}e^{-mT}(1-e^{-T})^{\mathsf{M}-1-m}\frac{1}{m+1}\\
 & =\frac{1}{\mathsf{M}}\sum_{m=0}^{\mathsf{M}-1}\binom{\mathsf{M}}{m+1}e^{-mT}(1-e^{-T})^{\mathsf{M}-1-m}\\
 & =\frac{e^{T}}{\mathsf{M}}(1-(1-e^{-T})^{\mathsf{M}}).
\end{align*}
Let $\mathsf{M}=e^{T-\delta}$. If $T\ge1$ and $\delta<0$, then the error probability is at least $1/12$. Hence we assume $\delta\ge0$. The correct probability is lower-bounded by
\begin{align}
\frac{e^{T}}{\mathsf{M}}(1-(1-e^{-T})^{\mathsf{M}})\ge\frac{e^{T}}{\mathsf{M}}(1-e^{-e^{-T}\mathsf{M}}) & =e^{\delta}(1-e^{-e^{-\delta}})\nonumber \\
 & \!\!\!\!\ge1-e^{-\delta}/2,\label{eq:correct_lb}
\end{align}
where the last inequality is due to $1-e^{-t}\ge t-t^{2}/2$. Hence, as long as $\mathsf{M}\le e^{T+\ln(2\epsilon)}$, we have $\delta\ge-\ln(2\epsilon)$, and the error probability is upper-bounded by $e^{-\delta}/2\le\epsilon$. For an upper bound on the correct probability, for $T\ge1$, 
\begin{align}
 & \frac{e^{T}}{\mathsf{M}}(1-(1-e^{-T})^{\mathsf{M}})\nonumber \\
 & \stackrel{(a)}{\le}\frac{e^{T}}{\mathsf{M}}\left(1-\exp\left(-\mathsf{M}e^{-T}(1+e^{-T})\right)\right)\nonumber \\
 & =e^{\delta}\left(1-\exp\left(-e^{-\delta}(1+e^{-T})\right)\right)\nonumber \\
 & \stackrel{(b)}{\le}1+e^{-T}-\frac{e^{-\delta}(1+e^{-T})^{2}}{2}\left(1-\frac{e^{-\delta}(1+e^{-T})}{3}\right)\nonumber \\
 & \le1-\frac{1}{2}e^{-\delta}(1-e^{-\delta}/2)+e^{-T},\label{eq:correct_ub}
\end{align}
where (a) is due to $\ln(1-t)\ge-t-t^{2}$ for $0<t\le1/2$, and (b) is due to $1-e^{-t}\le t-t^{2}/2+t^{3}/6$. The error exponent result directly follows from (\ref{eq:correct_lb}) and (\ref{eq:correct_ub}). From (\ref{eq:correct_ub}), if the error probability is at most $\epsilon$, then $e^{-\delta}/4\le\epsilon+e^{-T}$, and 
\begin{align*}
\epsilon & \ge\frac{e^{-\delta}}{2}(1-e^{-\delta}/2)-e^{-T}\ge\frac{e^{-\delta}}{2}(1-2(\epsilon+e^{-T}))-e^{-T},
\end{align*}
which gives $\delta\ge\ln(1/(2\epsilon+2e^{-T})-1)$. Hence, as long as $T\ge1$, the optimal $\mathsf{M}$ satisfies
\[
\exp\Big(T-\ln\frac{1}{2\epsilon}\Big)-1\le\mathsf{M}^{*}\le\exp\Big(T-\ln\Big(\frac{1}{2\epsilon+2e^{-T}}-1\Big)\Big).
\]
Therefore, $\ln\mathsf{M}^{*}=T-\ln(1/(2\epsilon))+O(\epsilon+e^{-T}/\epsilon)$. For a weaker bound, as long as $e^{-T}\le\epsilon\le1/12$, we have $\ln\mathsf{M}^{*}-(T-\ln(1/\epsilon))\in[0,2]$. Since $\mathsf{M}^{*}$ is non-decreasing in $T$, we have $\ln\mathsf{M}^{*}-\max\{T-\ln(1/\epsilon),0\}\in[0,2]$ whenever $0<\epsilon\le1/12$.  

\end{IEEEproof}
\begin{rem}
We may also study the identification problem \cite{ahlswede2002identification,salariseddigh2023deterministic,salariseddigh2024identification}, where the goal of the decoder is not to decode the message $K\in[\mathsf{M}]$, but rather to answer the question ``is $K=k_{0}$'' for a given $k_{0}$. For $(\mathsf{L},\alpha)$-OCPC with $\alpha<1$, $\mathsf{L}\ge\mathsf{M}$, the encoder can send $t\mapsto K$, and the decoder answers yes to the question if and only if $Y_{k_{0}}(T)\le cT$ for some decision boundary $c$ (change to $Y_{k_{0}}(T)\ge cT$ if $\alpha>1$). The false acceptance probability is $\mathbb{P}(Z_{2}\le cT)$ where $Z_{2}\sim\mathrm{Poi}(T)$, and the false rejection probability is $\mathbb{P}(Z_{1}>cT)$ where $Z_{1}\sim\mathrm{Poi}(\alpha T)$. For the perfect OCPC, setting $c=0$, we have a false acceptance probability $e^{-T}$ and no false rejection. The identification capacity is infinite if $\mathsf{L}=\infty$, $\alpha\neq1$.
\end{rem}

\section{Perfect Feedback and One-Cold Tree}

We now study the transmission of a source $S\in\mathcal{S}$, $S\sim P_{S}$ through the $(\mathsf{L},0)$-OCPC with perfect feedback. At time $t$, the encoder knows $Y_{i}(\tau)$ for $i\in[\mathsf{L}]$ and $\tau<t$, and can produce $x(t)$ accordingly. We will see that even when $\mathsf{L}$ is large enough, the encoder should sometimes group several values of $S$ into one band, instead of sending each value in its separate band. If the encoder initially sends $t\mapsto f(S)$ for a function $f:\mathcal{S}\to[\mathsf{L}]$, then there is no reason to change this strategy if no detection event occurs (no arrivals among $Y_{i}(\tau)$, $i\in f(\mathcal{S})$, $\tau<t$) since the posterior distribution of $S$ given $(Y_{i}(\tau))_{i\in[\mathsf{L}],\tau<t}$ is still $P_{S}$. However, if a detection event occurs in the $i$-th band $Y_{i}(\tau)$, then the encoder and the decoder know that $f(S)\neq i$, and the encoder should adjust $f$ to suit the posterior distribution of $S$. Hence, the strategy can be represented by a tree, where we start at the root, and move to the $i$-th child of the current node upon detection in the $i$-th band, until we reach a leaf which has a label indicating the decoded value of $S$. The tree is defined below.

\smallskip{}

\begin{defn}
An $\mathsf{L}$\emph{-ary one-cold tree} ($\mathsf{L}\in\{2,3,\ldots\}\cup\{\infty\}$) with alphabet $\mathcal{S}$ is a tree where each node has at most $\mathsf{L}$ children, each leaf is labelled by an element in $\mathcal{S}$, satisfying the property that $\mathcal{S}=\mathcal{S}(r)$ where $r$ is the root, and for each non-leaf node $u$ and each $s\in\mathcal{S}(u)$, there exists exactly one child $v$ of $u$ such that $s\notin\mathcal{S}(v)$, where $\mathcal{S}(u)$ denotes the set of labels appearing in the subtree with root $u$.
\end{defn}
\smallskip{}

Given an $\mathsf{L}$-ary one-cold tree, if the source is $S$ and the current node is $u$, the encoder finds the child $v$ of $u$ where $S\notin\mathcal{S}(v)$, and sends $t\mapsto i$ if $v$ is the $i$-th child of $u$. If a detection event occurs in the $j$-th band, we move to the $j$-th child of $u$ (where $S$ must appear in the subtree below this child). Note that a $2$-ary one-cold tree is precisely the usual notion of binary code tree for prefix codes. The optimal expected decoding time (the time we reach a leaf) can be given via dynamic programming in terms of the following quantity.

\smallskip{}

\begin{defn}
Given $S\sim P_{S}$ where $P_{S}$ is a discrete distribution with finite support $\mathcal{S}$ (i.e., $P_{S}(s)>0$ for $s\in\mathcal{S}$), its \emph{$\mathsf{L}$-ary one-cold entropy} is given recursively by $\bar{H}_{\mathsf{L}}(S):=0$ if $|\mathcal{S}|=1$, or otherwise
\begin{align*}
 & \bar{H}_{\mathsf{L}}(S)\;:=\\
 & \min_{f:\mathcal{S}\to[\mathsf{L}]:\,|f(\mathcal{S})|\ge2}\frac{C_{\mathsf{L}}+\sum_{i\in f(\mathcal{S})}\mathbb{P}(f(S)\neq i)\bar{H}_{\mathsf{L}}(S|f(S)\neq i)}{|f(\mathcal{S})|-1},
\end{align*}
where $C_{\mathsf{L}}:=-(\mathsf{L}-1)\ln(1-1/\mathsf{L})$ is the capacity of $(\mathsf{L},0)$-OCPC ($C=1$ if $\mathsf{L}=\infty$), and $\bar{H}_{\mathsf{L}}(S|f(S)\neq i)$ is the one-cold entropy of the conditional distribution of $S$ given $f(S)\neq i$.
\end{defn}
\smallskip{}

The optimal expected time is $\bar{H}_{\mathsf{L}}(S)/C_{\mathsf{L}}$. To see this, if the encoder initially sends $t\mapsto f(S)$, then the expected time of the next detection event among $|f(\mathcal{S})|$ bands is $1/(|f(\mathcal{S})|-1)$ since one band is blocked. The probability that the detection occurs in band $i$ is $\mathbb{P}(f(S)\neq i)/(|f(\mathcal{S})|-1)$. After detection, we switch to the optimal scheme for the conditional distribution of $S$ given $f(S)\neq i$. 

\begin{figure}
\begin{centering}
\includegraphics[scale=1.2]{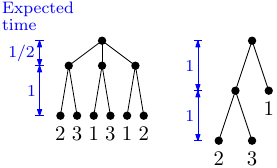}
\par\end{centering}
\caption{Two $3$-ary one-cold trees for $\mathcal{S}=\{1,2,3\}$ (assume $P_{S}(1)$ is the largest). The first tree corresponds to sending $t\protect\mapsto s$ initially, which takes an expected $1/2$ unit time to eliminate one value of $s$, and $1$ unit time to eliminate another value, resulting in $3/2$ unit time. The second tree corresponds to sending $t\protect\mapsto\min\{s,2\}$ initially (grouping $2,3$ together), using an expected $2-P_{S}(1)$ unit time in total. The optimal tree is one of these two trees.}\label{fig:tree}
\end{figure}

In particular, $\bar{H}_{2}(S)/\ln2$ is precisely the expected length of Huffman coding \cite{huffman1952method}. As a result, $H(S)\le\bar{H}_{2}(S)<H(S)+\ln2$. Also, for $S\sim\mathrm{Unif}([\mathsf{M}])$ and $\mathsf{L}=\infty$, the expected time is $\bar{H}_{\infty}(S)=\sum_{m=1}^{\mathsf{M}-1}m^{-1}$. For ternary $S$ with support $\mathcal{S}=\{1,2,3\}$, $\mathsf{L}\ge3$, the expected time can be computed as
\[
\bar{H}_{\mathsf{L}}(S)/C_{\mathsf{L}}=2-\max\big\{\max_{s}P_{S}(s),\,1/2\big\}.
\]
Refer to Figure \ref{fig:tree} for an illustration. Properties and bounds for general $\bar{H}_{\mathsf{L}}$ are left for future studies.

\smallskip{}

\section{Channel Simulation for OCPC}

Previous sections studied channel coding with OCPC, i.e., converting OCPC into bits. We now study channel simulation for OCPC, i.e., converting bits into OCPC. In channel simulation with unlimited common randomness \cite{bennett2002entanglement,cuff2013distributed,li2024channel}, the encoder and the decoder shares an arbitrary common randomness $W\sim P_{W}$. The encoder observes an input $X\in\mathcal{X}$, and sends $K=f(W,X)\in[\mathsf{M}]$ to the decoder. The decoder outputs $\hat{Y}=g(W,K)\in\mathcal{Y}$. The goal is to ensure that $\hat{Y}$ approximately follows a given conditional distribution $P_{Y|X}$, in the sense that the total variation distance $d_{\mathrm{TV}}(P_{\hat{Y}|X}(\cdot|x),P_{Y|X}(\cdot|x))\le\epsilon$ for every $x\in\mathcal{X}$, where $0<\epsilon<1$.  Let $\mathsf{M}_{\mathrm{sim}}^{*}(\mathsf{L},\alpha,T,\epsilon)$ be the smallest possible $\mathsf{M}$ such that there exists a coding scheme $(f,g)$ for simulating $P_{Y|X}$ given by the $(\mathsf{L},\alpha)$-OCPC, and let $\epsilon_{\mathrm{sim}}^{*}(\mathsf{L},\alpha,T,\mathsf{M}):=\inf\{\epsilon\ge0:\mathsf{M}_{\mathrm{sim}}^{*}(\mathsf{L},\alpha,T,\epsilon)\le\mathsf{M}\}$ be the optimal total variation distance for a given $\mathsf{M}$. We now give an achievability result for the perfect OCPC.

\smallskip{}

\begin{thm}
\label{thm:finite_L-1-1}For the perfect OCPC, we have
\begin{align*}
\epsilon_{\mathrm{sim}}^{*}(\infty,0,T,\mathsf{M}) & \le(1-e^{-T})^{\mathsf{M}}\le\exp(-e^{-T}\mathsf{M}).
\end{align*}
As a result, 
\[
\ln\mathsf{M}_{\mathrm{sim}}^{*}(\infty,0,T,\epsilon)\le T+\ln\ln(1/\epsilon)+\ln2.
\]
\end{thm}
\smallskip{}

\begin{IEEEproof}
The common randomness consists of i.i.d. Poisson processes $\tilde{Y}_{i}^{(k)}(t)$ with rate $1$ for $i\in[\mathsf{L}]$, $k\in[\mathsf{M}]$. Given input $X:[0,T]\to[\mathsf{L}]$, the encoder uses rejection sampling to find the smallest $K\in[\mathsf{M}]$ such that for every $i$, $\tilde{Y}_{i}^{(K)}(t)$ contains no points at time $t$ where $X(t)=i$ (which is the requirement of the output of the perfect OCPC). The decoder simply outputs $\hat{Y}_{i}(t)=\tilde{Y}_{i}^{(K)}(t)$. The probability that the encoder fails to find $K$ is $(1-e^{-T})^{\mathsf{M}}\le\exp(-e^{-T}\mathsf{M})$. We obtain the desired results. 
\end{IEEEproof}
\smallskip{}

For general OCPC, we have the following bound.

\smallskip{}

\begin{thm}
\label{thm:finite_L-1}For the $(\mathsf{L},\alpha)$-OCPC, we have
\begin{align*}
 & \epsilon_{\mathrm{sim}}^{*}(\mathsf{L},\alpha,T,\mathsf{M})\;\le\\
 & \,\mathbb{E}\big[\big(1-\big(1+\alpha^{Z_{1}}(1+(\alpha-1)/\mathsf{L})^{-Z_{1}-Z_{2}}\big)^{-1}\big)^{\mathsf{M}}\big],
\end{align*}
where $Z_{1}\sim\mathrm{Poi}(\alpha T)$ is independent of $Z_{2}\sim\mathrm{Poi}((\mathsf{L}-1)T)$ (replace the $(1+(\alpha-1)/\mathsf{L})^{-Z_{1}-Z_{2}}$ term by $e^{(1-\alpha)T}$ if $\mathsf{L}=\infty$). As a result, for any fixed $0<\epsilon<1$, for $T\to\infty$,
\[
\ln\mathsf{M}_{\mathrm{sim}}^{*}(\mathsf{L},\alpha,T,\epsilon)\le CT+\sqrt{VT}Q^{-1}(\epsilon)+O(\log T),
\]
where $C$, $V$ are the capacity and dispersion in Theorem \ref{thm:capacity}.
\end{thm}
\smallskip{}

\begin{IEEEproof}
We use the Poisson functional representation \cite{li2021unified}. The common randomness consists of i.i.d. Poisson processes $\tilde{Y}_{i}^{(k)}(t)$ with rate $\gamma=1+(\alpha-1)/\mathsf{L}$ for $i\in[\mathsf{L}]$, $k\in\mathbb{N}$, as well as a Poisson point process $(B_{k})_{k\in\mathbb{N}}$ (i.e., $B_{1},B_{2}-B_{1},\ldots\stackrel{\mathrm{iid}}{\sim}\mathrm{Exp}(1)$). Given input $X:[0,T]\to[\mathsf{L}]$, the encoder sends 
\[
K=\mathrm{argmin}_{k}B_{k}\Big(\frac{\mathrm{d}P_{(Y_{i}(t))_{i,t}|X}(\cdot|X)}{\mathrm{d}P_{(\tilde{Y}_{i}(t))_{i,t}}}((\tilde{Y}_{i}^{(k)}(t))_{i,t})\Big)^{-1},
\]
where $P_{(Y_{i}(t))_{i,t}|X}$ is the conditional distribution given by the OCPC, and $P_{(\tilde{Y}_{i}(t))_{i,t}}$ is the distribution of i.i.d. Poisson processes with rate $\gamma$. If $K$ has no upper bound, then the OCPC can be simulated with zero TV distance. Nevertheless, since $K\le\mathsf{M}$, we have an error when $K>\mathsf{M}$, which has a probability upper-bounded by the generalized Poisson matching lemma \cite[Lemma 3]{li2021unified}, giving the desired result. The bound on $\mathsf{M}_{\mathrm{sim}}^{*}$ follows from Gaussian approximation.
\end{IEEEproof}
\smallskip{}

\begin{rem}
We may also study variable-length channel simulation \cite{harsha2010communication,sfrl_trans}, where the encoder sends $K\in\mathcal{C}$ where $\mathcal{C}\subseteq\{0,1\}^{*}$ is a prefix code. By the strong functional representation lemma \cite{sfrl_trans,li2024pointwise,li2024channel}, the expected length of $K$ can be upper-bounded by $C+\log_{2}(C+2)+3$, where $C$ is the capacity of the OCPC.
\end{rem}
\smallskip{}

\section{Acknowledgement}

This work was partially supported by two grants from the Research Grants Council of the Hong Kong Special Administrative Region, China {[}Project No.s: CUHK 24205621 (ECS), CUHK 14209823 (GRF){]}.

\smallskip{}

\bibliographystyle{IEEEtran}
\bibliography{ref}

\end{document}